\documentclass[11pt,preprint,flushrt]{aastex}

\def\eqq#1{Equation~(\ref{#1})}
\newcommand\etal{{\it et al.\/}}
\newcommand\eg{{\it e.g.\ }}
\newcommand\ie{{\it i.e.\ }}

\newcommand{\bfk}{\mbox{\boldmath $\kappa$}}

\newcommand{\bft}{\mbox{\boldmath $\theta$}}

\begin{document}

\slugcomment{CVS \$Revision: 2.6 $ $ \$Date: 2005/09/13 18:57:28 $ $
Accepted to ApJ}

\title{Metric Tests for Curvature from Weak Lensing and Baryon
  Acoustic Oscillations} 

\author{G. Bernstein}
\affil{Dept. of Physics and Astronomy, University of Pennsylvania,
Philadelphia, PA 19104}
\email{garyb@physics.upenn.edu}

\begin{abstract}
We describe a practical measurement of the curvature of the
Universe which, unlike current constraints, relies purely on the
properties of the Robertson-Walker metric rather than any assumed
model for the dynamics and content of the Universe.
The observable quantity is
the cross-correlation between foreground mass and
gravitational shear of background galaxies, which depends upon the
angular diameter distances $d_A(z_\ell)$, $d_A(z_s)$, and
$d_A(z_s,z_\ell)$ on the degenerate triangle formed by observer,
source, and lens.  In a flat Universe, 
$d_A(z_\ell,z_s) = d_A(z_s)-d_A(z_\ell)$, but
in curved Universes an additional term $\propto\Omega_k$
appears and alters the lensing observables even if $d_A(z)$ is fixed.
We describe a method whereby weak lensing data may be used to
solve simultaneously for $d_A$  and the curvature.
This method is completely
insensitive to the equation of state of the 
contents of the Universe, or amendments to General Relativity
that alter the gravitational deflection of light or the growth of
structure. The curvature estimate is also independent of biases
in the photometric redshift scale.
This measurement is
shown to be subject to a degeneracy among $d_A$, $\Omega_k$ and the
galaxy bias factors that
may be broken by using the same imaging data to
measure the angular scale of baryon acoustic oscillations.
Simplified estimates of the
accuracy attainable by this method indicate that ambitious
weak-lensing $+$ baryon-oscillation surveys would measure $\Omega_k$
to an accuracy 
$\approx0.04 f_{\rm sky}^{-1/2} (\sigma_{\ln z}/0.04)^{1/2}$, where
$\sigma_{\ln z}$ is the photometric redshift error.
The Fisher-matrix formalism developed here is also useful for 
predicting bounds on curvature and other characteristics of parametric
dark-energy models.  We forecast some representative error levels and
compare to other analyses of the weak lensing cross-correlation
method.  We find both curvature and parametric constraints to be
surprisingly insensitive to the systematic shear calibration errors.
\end{abstract}


\section{Metric Measurements of Curvature}
The most robust prediction of inflation theories is that the radius of
curvature of the present Universe should be very large,
\ie\ $\Omega_k=0$ to high accuracy.  The WMAP$+$supernovae$+H_0$ best
fit of 
$\Omega_k=-0.02\pm0.02$ \citep{Spergel} is therefore taken as a vindication of
inflation, and a great majority of current work, such as
investigations of the properties of dark energy, take $\Omega_k=0$ as
{\it a priori} truth.  It is important to realize, however, that WMAP
does not measure curvature in any direct geometric way.
Constraints on curvature derive primarily from a measurement of the
angular diameter distance to recombination, $D_A(z_{\rm rec})$, which
depends upon curvature but also upon $\Omega_m$ and models for dark
energy or other constituents of the Universe.  Hence the curvature
results are dependent upon the $\Lambda$CDM (or other)
model for dark energy assumed in the analysis.  
Our ignorance of the
dark energy phenomenon limits our ability to test for flatness or to
look for small finite $\Omega_k$ that may be predicted in
variants of inflation \citep{Uzan} or in Universes with non-trivial
topology \citep{Luminet}.  

Our ignorance of the true curvature will, conversely, foil attempts to
characterize dark energy properties with Type Ia supernovae (or other
standard candles), which measure $D_L(z)$ at lower redshifts.  If
the Tolman surface-brightness relation holds, then 
we have $D_A=D_L(1+z)^{-2}$. The
proper-angular-diameter distance in a Universe with
Robertson-Walker metric is
\begin{eqnarray}
D_A(z) & = & (1+z)^{-1} S_k\left[r(z)\right], \\
S_k(r) & \equiv & 
\left\{
\begin{array}{cl}
R_0 \sin(r/R_0) & k=+1 \\
r & k=0 \\
R_0 \sinh(r/R_0) & k=-1 
\end{array}
\right.  \\
r(z) & = & c \int_0^z dz H^{-1}(z).
\end{eqnarray}
Here $r$ is the comoving radial distance, $R_0$ is
the radius of 
curvature of the Universe, and $H(z)=\dot a /a$.  These equations
follow purely from the RW metric.  If standard-candle and CMB
observations were to give us perfect knowledge of $D_A(z)$, then for
any $R_0$ and $k$ there is a solution 
\begin{equation}
{c \over H(z)} = {d \over dz} S_k^{-1}\left[(1+z)D_A(z)\right].
\end{equation}
that will exactly reproduce the data.  If the Friedmann equations hold, then
\begin{eqnarray}
H(z) &=&  H_0 \left[\Omega_m(1+z)^3 + \Omega_k(1+z)^2 +
  (1-\Omega_m-\Omega_k)f_X(z)\right]^{1/2}, \\
\Omega_k & \equiv & {-kc^2 \over H_0^2R_0^2}.
\end{eqnarray}
Here $f_X(z)$ describes the evolution of $\rho_X$, an additional
dark-energy component.  Hence even with perfect knowledge of $D_A(z)$,
there is always some dark energy behavior $f_X$ which reproduces that data
for {\it any} choice of curvature.  The degeneracy between curvature
and dynamics is well known, \eg\ \citet{Weinberg}
demonstrates 
that even complete knowledge of $D_A(z)$ cannot constrain the
curvature in the absence of a dynamical model for the expansion.
Constraints on curvature from
$D_A$ data arise solely because of our preferences for forms
of $f_X$ that arise from certain equations of state.  Allowing
alterations to the Friedmann equations adds additional degeneracy.

\begin{figure}[t]
\epsscale{0.7}
\plotone{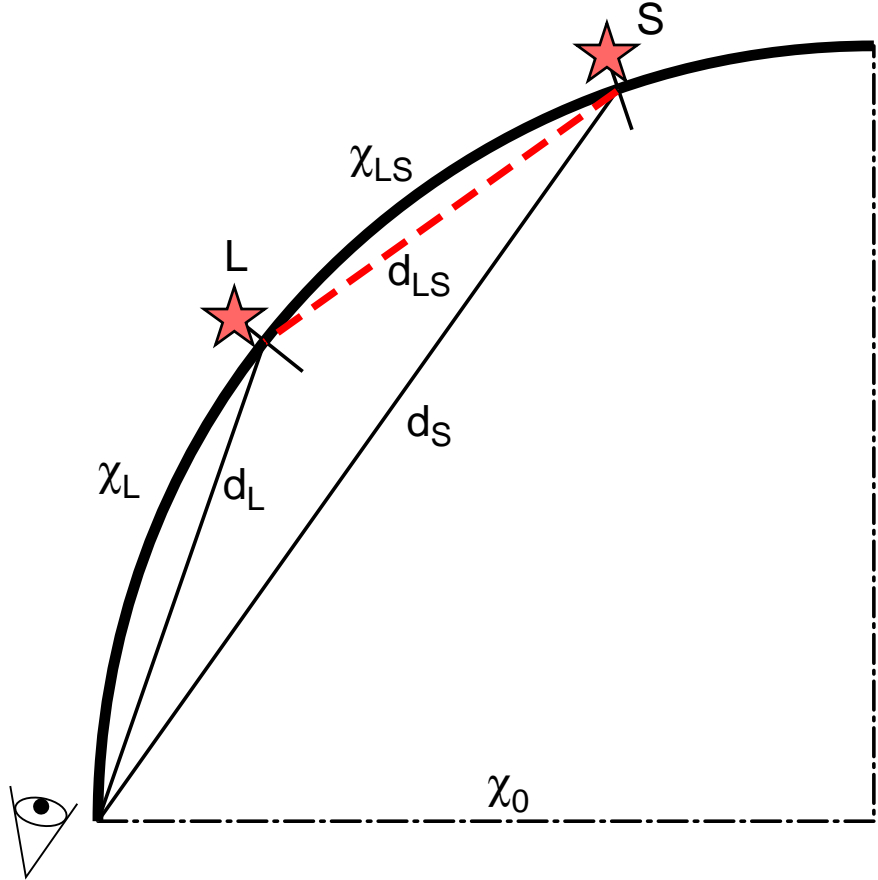}
\caption[]{
\small Illustration of curvature determination: consider an observer
on a section of the closed Universe, represented by the bold line.
The conformal distances $\chi_L$ and $\chi_S$ to two sources are not
directly observable; only the angular-diameter distances are, $d_L$ and
$d_S$ represented here by the chord lengths.  The radius of curvature
$\chi_0$ 
is indeterminate, but a measure of the distance $d_{LS}$ (dashed line)
allows determination of the curvature.
}
\label{chords}
\end{figure}

We can break the degeneracy between curvature and $f_X$ without
making dynamical assumptions if we apply
the metric to some line segment that does not originate at
$z=0$, essentially forming a cosmological-scale triangle.
For example if a
gravitational lens at $z_\ell$ bends a light ray by angle $\alpha$,
then the apparent deflection of a source at $z_s$ observed from $z=0$ obeys
\begin{equation}
\label{dtheta}
\delta\theta =  \alpha {D_A(z_\ell, z_s) \over D_A(z_s)},
\end{equation}
where $D_A(z_\ell, z_s)$ is the proper angular-diameter distance at
  $z_s$ as viewed from $z_\ell$.  We introduce a dimensionless comoving angular
  diameter distance
\begin{eqnarray}
d(z_1,z_2) & \equiv & {H_0 \over c} (1+z_2)D_A(z_1,z_2) \\
 & = & S(\chi_2 - \chi_1), \\
\chi & \equiv & {H_0 \over c} r, \\
d(z) & \equiv & d(0,z) \\
S(\chi) & = & 
\left\{
\begin{array}{cl}
\chi_0 \sin(\chi/\chi_0) & k=+1 \\
\chi & k=0 \\
\chi_0 \sinh(\chi/\chi_0) & k=-1 
\end{array}
\right.
\end{eqnarray}
Now $\chi_0$ is the radius of curvature in Hubble lengths.  Using the
difference rule for $S(\chi)$ we obtain
\begin{eqnarray}
d(z_\ell,z_s) & = & S(\chi_s - \chi_\ell) \\
 & = & S(\chi_s)C(\chi_\ell) - S(\chi_\ell)C(\chi_s) \\
C(\chi) & \equiv & 
\left\{
\begin{array}{cl}
\cos(\chi/\chi_0) & k=+1 \\
1 & k=0 \\
\cosh(\chi/\chi_0) & k=-1 
\end{array}
\right. \\
\Rightarrow \quad
d(z_\ell,z_s) & = & S(\chi_s)\left[1 + {\Omega_k \over 2}
  S^2(\chi_\ell)\right]
 - S(\chi_\ell)\left[1 + {\Omega_k \over 2} S^2(\chi_s)\right] +
   O(\Omega_k^2) \\
\label{dls}
 & = & (d_s - d_\ell)(1 - \Omega_kd_\ell d_s /2) + O(\Omega_k^2).
\end{eqnarray}
So while the
$d(z)$ function itself has a curvature-dark energy degeneracy, the
lensing strength depends purely on $\Omega_k$ once $d(z)$ is known,
\eg\ from supernova measurements.  This $\Omega_k$ dependence is well
known, \eg\ 
Equation~(13.70) in \citet{Peebles}. Figure~\ref{chords} illustrates
how the curvature of the Universe is indeterminate when distances
from $z=0$ are the only observables, whereas a measure of $d_{\ell s}$
determines curvature.  \citet{EVL88} remarks that
the presence of this differential distance in gravitational lensing
equations could provide useful cosmological tests, but to our
knowledge no practical implementation of this---or any other---metric
curvature test has been proposed.  

We propose that $\Omega_k$ be determined by measuring the
gravitational lensing shear $\gamma$ of background galaxies---\ie\ the
gradient of 
$\delta\theta$ in \eqq{dtheta}---and measuring the amplitude of the
contribution from the $O(\Omega_k)$ term in \eqq{dls}.
A foolishly optimistic estimate of the accuracy available
on $\Omega_k$ can be obtained by assuming that the deflection angles
$\alpha$ are known {\it a priori} at all redshifts and positions on
the sky, and that the angular diameter distance $d(z)$ has been
measured as well.  Since
$d_\ell$ and $d_s$ are in the range 0.5--1.5 at relevant
distances for weak lensing in the $\Lambda$CDM fiducial model, the
lensing shear scales as $\approx(1-\Omega_k/2)$, so we would find
that $\delta\Omega_k \approx 2 \delta\gamma / \gamma$.  The typical
shear amplitude $\gamma$ over a line of sight is $\approx0.02$ at
cosmological distances.  The uncertainty in shear is $\approx
0.3/\sqrt N_g$, where $N_g$ is the number of galaxies with
well-measured shapes.  A space-based survey can measure 100 galaxies
per square arcminute, which leads to
$\delta \Omega_k \sim 0.0002/\sqrt f_{\rm sky}$, far
better than the current model-dependent constraints.  This estimate is
unrealistic, however, because neither the distance functions nor the
deflections will be known.  Below we will investigate a simultaneous
solution for the angular-diameter distances, deflection strengths, and
$\Omega_k$ from cross-correlation of lensing shear with foreground
galaxy distributions, and estimate the resultant degradation in the
curvature constraint.

If we execute the WL curvature measurement by dividing our galaxy
sample into redshift bins, then we note that we do not actually need
to know the redshift of each bin; it is only the
distance $d$ of each bin that enters into the calculation of the shear.
Hence biases in photometric redshifts do not affect the curvature
measurement. 
It is essential, however, to insure that the galaxies in a bin are at
a common distance.

In the remainder of this section we examine metric
tests for curvature which might be possible with other cosmological
observables besides weak lensing.  In \S2 we develop the methodology
for using weak lensing cross-correlations to solve simultaneously for
the curvature, 
distance relationships, and deflector properties.  The implementation
is significantly more involved and subtle than the simple idea that
curvature is manifested in $d_{\ell s}$.  \S3 applies the formalism to
forecast curvature constraints from feasible surveys, and extends the
formalism to include treatment of the likely dominant systematic
error.  \S4 investigates the constraints on parametric models of dark
energy ($w_0$,$w_a$) that can be derived with the curvature left free
to vary, and compares the present results to some previous work on
weak lensing cross-correlations.  \S5 summarizes and concludes.

\subsection{Metric Curvature from Other Cosmological Measurements}
Similarly metric determinations of $\Omega_k$ are in principle
derivable 
from other cosmological measurements.  The most promising is the
detection of the recombination acoustic horizon scale, which is known
in comoving physical units from the CMB, in the power spectrum of
galaxies, recently demonstrated by \citet{SDSSBAO}.
The transverse baryon acoustic oscillation (BAO) test measures the
angular scale subtended by this 
standard ruler, and hence measures $(1+z)D_A(z)\propto h^{-1}d(z)$.
The transverse BAO method does not by itself break the
curvature-dark energy degeneracy, but may serve as a
source of high-reliability constraints on $d(z)$ to combine with weak
lensing information.  Again for a given galaxy subsample, the redshift
itself is unimportant, just the observable $d$. 

With sufficient redshift resolution, the acoustic scale may be
identified along the line of sight, yielding knowledge of 
\begin{equation}
{dr \over dz} = {c \over H_0} d^\prime(z) (1 - \Omega_kd^2/2)
+O(\Omega_k^2).
\end{equation}
Hence the comparison of line-of-sight to transverse BAO
scales can yield $\Omega_k$ directly, but requires one to either
integrate the line-of-sight $H(z)$ data or differentiate the
transverse $d(z)$ data.
Doing either operation in the presence of noise adds substantial
difficulties to making a high-accuracy comparison, unless one assumes
a parametric form for $d(z)$.

Counts of galaxy clusters are sensitive to $dV/d\Omega\,dz \propto
h^{-3} d^2 d^\prime(z) (1 - \Omega_kd^2/2)$.  Volume-element data alone can
not measure $\Omega_k$, but could potentially do so in combination
with high-precision measures of $d(z)$ {\em and} its derivative.  Cluster
counts are also, however, dependent upon the growth of structure, so
do not offer the model-independent cosmographic information of weak
lensing or baryon oscillations.

Stellar evolution has been proposed
as a standard chronometer \citep{Raul}, which could determine $H(z)$
and hence the curvature if combined
with measures of $d$.  To contribute to the metric measurement of
curvature, accuracies of a few percent on $H(z)$ are required, and
further investigation into the complexities of stellar and galactic
evolution are needed in order to determine whether robust constraints
at this accuracy are possible.

Strong lensing may also be used to discern the behavior of
$d(z_\ell,z_s)/d(z_s)$, if one is presented with a lensing system with
multiply-imaged sources at a variety of (known) background redshifts.
Such use of multiple arcs around clusters has been investigated in the
context of constraining dark energy \citep{LP98, GFM, GKS,
S02,Soucail}, and would be equally applicable
to determination of curvature, but also equally susceptible to the
small uncertainties in the mass profiles of clusters.

\section{Cosmographic Methodology}
The metric curvature determination is clearly
an extension of the cross-correlation cosmography (CCC) technique
proposed by \citet{JT}: one
identifies rich foreground clusters and measures the
dependence of the induced background shear upon $z_s$ of the source
galaxies.  The mass of the cluster(s) may be unknown, but the {\em ratio} of
any two background shears is a purely cosmographic function.
\citet{SK04} combine this approach with other dark energy probes.
\citet{BJ04}[BJ04] generalize the method to a cross-correlation between
foreground estimated-mass distributions and background shear
patterns.  In this case the cluster-mass uncertainty is replaced by an
unknown bias factor for each foreground mass shell, over which we
marginalize.   
\citet{Zhang}[ZHS] give a more elegant analysis in which the galaxy-shear
cross-correlations are considered simultaneously with the shear-shear
correlations in a single Fisher matrix.  \citet{HJ04} also include the
galaxy-galaxy correlations in the same Fisher matrix, and make use of
a parametric model of bias based on the halo model of galaxy
distributions.  All of these analyses differ substantially in 
their analytic approaches, underlying assumptions, and estimated
constraints.  A full comparison is beyond the scope of this paper, but
we will make some brief comments below.  Here we attempt to recast the
problem in a manner suited to the metric curvature constraint, but
also amenable to studying the parametric dark-energy constraints
investigated by these authors.

The shear induced in direction \bft\ on sources at $z_s$ by mass in
the foreground can be decomposed into $E$- and $B$-mode 
components.  In the weak-lensing limit, the former is
straightforwardly related to the lensing 
convergence $\kappa$ and the latter should vanish.  Hence we will
consider the shape information to come in the form of the convergence
field $\kappa$, which depends upon direction and source redshift as
\begin{equation}
\label{kappa1}
\kappa(\bft, z_s) = {3 \Omega_m \over 2 }
  \int_0^{z_s} dz_\ell {d_\ell(1+z_\ell)H_0\over H(z_\ell)} m({\bft},z_\ell)
{d_s - d_\ell \over d_s}(1-\Omega_kd_sd_\ell/2),
\end{equation}
where $m$ is the overdensity $\delta\rho/\bar\rho$ of the total mass.

We also
idealize the sky as consisting of $K$ discrete pencil-beams of solid
angle $\delta\Omega$, with the true mass density $m(\bft_i)$
being statistically independent of the density in any other beam.
We can choose $\delta\Omega$ to correspond to a correlation length of
the true continuum shear field.  We will assume
$\delta\Omega=0.25\,{\rm arcmin}^2$; the results depend only weakly on
this choice.  We
could equivalently decompose the shear and mass distributions into
spherical harmonics, in which case our choice of $\delta\Omega$
becomes equivalent to a maximum multipole $\ell\approx2\times10^4$.

The weak lensing equation (\ref{kappa1}) can be written
\begin{eqnarray}
\label{gcontinuum}
\kappa(s,\bft_i) & = & \int d\ell\,G(s,\ell)m(\ell,\bft_i) k(\ell), \\
k(\ell) & \equiv & {3\Omega_m\over 2} \ell (1+z_\ell) (1-\Omega_k \ell^2/2)  \\
G(s,\ell) & = & \left(1-{\ell \over s}\right) (1-\Omega_k s \ell/2)
\Theta(s-\ell),
\end{eqnarray}
where $\Theta$ is the step function, and we have reparameterized the source- and
lens-plane distances by $s=d(z_s)$, $\ell=d(z_\ell)$.  This equation
assumes the validity of General Relativity in two respects: first,
that light follows the geodesic equation for the metric, and second,
that we have the usual Poisson equation.  The second assumption can be
dropped in our quest for a purely metric test of curvature.  In our
formalism a change of the Poisson equation would be accommodated by
generalizing $k_\ell m(\ell,\bft)$ to mean the convergence caused by
the mass at distance $\ell$, even if this means that $m$ is no
longer the unaltered mass distribution.

We discretize the analysis by
considering the galaxies to be located on a series of $J$ shells, in
order to facilitate the construction of Fisher matrices below.  The
finite thickness of the shells complicates the analysis (ZHS)
and in fact it is possible to draw erroneous conclusions from a
discrete analysis, as discussed by \citet{Albert}.  
Below we will attempt to return the discrete formulation to the
continuum limit and discuss these issues.  

The measured shear at each source plane becomes, with a noise term now
added, 
\begin{eqnarray}
\label{gdiscrete}
\kappa_s(\bft_i) & = & \sum_\ell G_{s\ell} m_\ell(\bft_i) k_\ell +
(\delta\kappa)_s(\bft_i), \\ 
\label{gls}
G_{s\ell} & = & \left\{
\begin{array}{cl}
{d_s - d_\ell \over d_s}(1-\Omega_kd_sd_\ell/2) & \ell<s \\
0 & \ell \ge s
\end{array}
\right . \\
\label{vargamma}
{\rm Var}(\delta\kappa_s) & = & {\sigma_\gamma \over
  \sqrt{n_s\,\delta\Omega}}, \\
k_\ell & \equiv & {3\Omega_m\over 2} d_\ell (1+z_\ell)\Delta \chi =
 {3\Omega_m\over 2} d_\ell (1+z_\ell)(1-\Omega_kd_\ell^2/2)\Delta d_\ell.
\end{eqnarray}
Now $m_\ell(\bft)$ is the mass overdensity averaged through the shell
of width $\Delta \chi$; $k_\ell$ is a factor that converts this
overdensity into the convergence measured on a lens plane at infinity.
The shear measurement noise is determined by the areal density $n_s$
of shape-measurable galaxies on shell $s$, and
$\sigma_\gamma\approx 0.3$ \citep{wlerrs}.

The true mass fluctuation $m_\ell(\bft)$ is not directly observable.  But
we assume that we deduce from the galaxy distribution some estimate of
the mass overdensity $g_\ell(\bft)$ that is correlated with the
true distribution.  This proxy field need not be the
galaxy density itself, but rather the result of any algorithm applied
to the observed galaxy field, \eg\ involving the
assignment of halos to galaxies, groups, and clusters.  The
correlation coefficient of the true and proxy mass fields is
\begin{equation}
{\langle m_i g_j \rangle \over
\sqrt { \langle m_i^2 \rangle \langle g_j^2\rangle } } = \delta_{ij}
r_i,
\end{equation}
where the angle brackets denote averaging over direction \bft.  The
$r_i$ are not known but are assumed positive.  We assume that the
shells are thick enough to be uncorrelated.  We 
also define for each shell an unknown bias via
\begin{equation}
 \langle m_i g_i \rangle = B_i \langle g_i^2 \rangle.
\end{equation}
Note that this is the bias of the {\em true} mass with respect to the
{\em estimated} mass.  The more conventional bias $b$ of the estimated
(galaxy) mass to the true mass is, for Gaussian-distributed $g$ and
$m$, related to this via 
\begin{equation}
B_i = {r_i^2 \over b_i}.
\end{equation}
The shear noise $\delta\kappa$ is assumed to have correlations with
neither the true nor the proxy mass distribution. 

With these definitions we could
extract curvature information from our observed shear data as
follows: we cross-correlate each convergence field $\kappa_s$ with each
foreground proxy field $g_\ell$ to give an observable quantity
$X_{s\ell}$.  From \eqq{gdiscrete} the expectation values are
\begin{equation}
\label{expect}
\langle X_{s\ell} \rangle = \sum_{\ell^\prime}G_{s\ell^\prime}
k_{\ell^\prime}
\langle g_\ell m_{\ell^\prime} \rangle = G_{s\ell}B_\ell k_\ell \langle
g_\ell^2 \rangle.
\end{equation}
There are $J(J-1)/2$ non-zero observable cross-correlations, which
are statistically independent for any significant sky coverage. These 
$X_{s\ell}$ are determined by the $J$ factors $(B_\ell
k_\ell\langle g_\ell^2 \rangle)$, plus the
matrix ${\bf G}$ that is defined by $J$ distances $d_i$ and the single
$\Omega_k$.  Hence there are $2J+1$ free parameters, so we might expect 
an unambiguous fit to the data for $J\ge5$.  After marginalization over the
nuisance parameters $B_i$ and $d_i$ (which carry much information
about dark energy and galaxy formation!) one obtains constraints on
$\Omega_k$.  

Unfortunately the simultaneous solution for $\Omega_k$, ${\bf B}$, and
${\bf d}$ from the system of equations (\ref{expect}) has three
degeneracies in the neighborhood of $\Omega_k=0$.  The transformations
\begin{eqnarray}
\label{degen1}
 & d_i \rightarrow d_i(1+\alpha_0) & \\
\label{degen2}
& d_i \rightarrow d_i/(1-\alpha_1d_i), \quad
 (B_i k_i)\rightarrow (B_i k_i)(1-\alpha_1d_i) & \\
\label{degen3} 
& d_i \rightarrow d_i/(1-\alpha_2d_i^2), \quad
(B_i k_i)\rightarrow (B_i k_i)(1-\alpha_2d_i^2), \quad \Omega_k\rightarrow
 \Omega_k+2\alpha_2 &
\end{eqnarray}
each leave the observable shear unchanged to first order in $\Omega_k$
and the free parameters $\alpha_i$.  The first degeneracy is a simple
scaling of the $d_i$, which is unsurprising as the lensing observables
depend only upon distance ratios.  The second degeneracy is similarly
benign as it leaves the solution for $\Omega_k$ unchanged.  But the
third degeneracy leaves $\Omega_k$ indeterminate.  The degeneracy can
be broken if we assume some functional form for $d(z)$, but our goal
is a model-independent constraint.  The same imaging survey that is
used to generate the shear measurement and the foreground galaxy maps
can be used to measure the transverse baryon acoustic horizon scale,
if the photometric redshift accuracy $\sigma_{\ln z}$ is sufficiently
good.  This will produce a statistically independent measure of $d_i$
at each shell, which we may use to break the curvature degeneracy that
remains in the lensing cross-correlations.

There is a more subtle problem with a solution for cosmology via
\eqq{expect}.  As noted by \citet{Albert}, the solution offers no
discriminatory power on cosmology if we allow the $B_i$ to be
completely free, because the matrix $G_{s\ell}$ must be invertible
as we approach the continuum limit.  Looking explicitly at the
continuum limit in \eqq{gcontinuum}, Stebbins shows that there must
always be some function $G^{-1}(\ell, s)$ such that
\begin{equation}
m(\ell)k(\ell) = \int ds\,G^{-1}(\ell, s) \kappa(s).
\end{equation}
In fact we can construct this function explicitly to first order in
$\Omega_k$: 
\begin{eqnarray}
\label{ginv}
G^{-1}(\ell, s) & = & \left(1 + {\Omega_k \over
  2}\ell^2\right)s\delta^{\prime\prime}(s - \ell)
 + \Omega_k\ell s \delta^\prime(s-\ell) \\
\Rightarrow \qquad 
m(\ell) k(\ell) & = & \left(1 + {\Omega_k \over
  2}\ell^2\right)\left. {d^2(s\kappa) \over ds^2}\right|_\ell
 + \Omega_k\ell s\left. {d(s\kappa) \over ds}\right|_\ell.
\end{eqnarray}
Here $\delta$ is the Dirac function.  Surprisingly this inversion of
the lensing cross-correlation is local.  Given this solution,
we simply set $B(\ell) = m(\ell) / g(\ell)$ and get an
exact solution on this line of sight, regardless of our choice of
$\Omega_k$ and $d(z)$ functions.  There is hence no possibility of
inference of cosmological parameters along this line of sight as we
reach the continuum limit.  Similarly a solution of the
cross-correlation \eqq{expect} is possible for any choice of
cosmology, once we reach a large number of lens and source planes, as
$G$ is always invertible to give a valid ${\bf B}$.

An exit from this conundrum is the realization that giving complete
freedom to the bias function $B(\ell)$ is tantamount to discarding our
initial presumption that the proxy field is correlated with the true
mass distribution.  Imagine slicing the Universe so finely as to
assign each proton its own bias value with respect to the dark
matter.  With this freedom we could clearly produce any macroscopic mass
distribution we wish, including one that has no
correlation at all with the galaxy-scale baryon distribution.  We must
therefore incorporate into the analysis some criteria for the
coherence of the bias parameter in order to reflect our underlying
assumption that mass has some correlation with the proxy field.

We note, for example, that the continuum-limit inversion formula
\eqq{ginv} will 
produce a divergent solution for $m(\ell)$ in the presence of any
finite amount of shot noise on the lensing field $\kappa(s)$.  The
cross-correlation between estimated $m(\ell)$ and $g(\ell)$ can be
estimated from the multiple lines of sight in the survey, and will be
driven to zero in this continuum limit.  If the likelihood function
for our joint analysis of the $\kappa$ and $g$ fields requires them to
be correlated Gaussian variables, then the rapidly varying components of the
inversion solution for $m$ will
be suppressed in a maximum-likelihood solution, and it becomes
possible to discriminate cosmologies.

We do not claim yet to have a rigorous model-independent
method to approach the continuum limit.
Below we produce a likelihood function for the shear and proxy fields
that incorporates their presumed correlation on scales of a shell
thickness.  Our practical approach 
will be to see how the uncertainties in our cosmological parameters
scale as the number of shells increases; in fact we see no significant
changes in any of the results below as we decrease the shell size as
far as $\Delta z \approx 0.02$.  Another practical approach
that we will take is to include a regularization condition on $B(\ell)$
to lower the probability of solutions that vary rapidly on short
redshift (and hence time) scales, in accordance with our physical
intuition; this too has no significant impact upon our forecasts.  

A potentially more enlightening approach to the cross-correlation
problem may be to parameterize the functions $d(z)$ and $B(z)$ by
expansions in orthonormal function sets, such as Fourier modes or
Legendre polynomials, rather than assume a stepwise variation and/or
discrete galaxy distributions.

\subsection{Likelihood Function}
The accuracy of constraints on $\Omega_k$ can be estimated by the
Fisher matrix methodology.  We produce a Fisher matrix for a single
line of sight, and then multiply by $K$ since we have assumed all
lines of sight to be independent.  On each line of sight there are
three $J$-dimensional vectors: the measured convergence \bfk; the true mass
distribution ${\bf m}$; and the proxy mass estimate ${\bf g}$.  We
need a likelihood function $L(\bfk, {\bf m}, {\bf g})$, which we then
must marginalize over the unobservable ${\bf m}$.  This probability
can be expressed as 
\begin{equation}
L(\bfk, {\bf m}, {\bf g}) = L(\bfk | {\bf m}) 
L({\bf m}, {\bf g}) .
\end{equation}
Since the shear measurements arise from the sum of many individual
galaxy shape measurements, the shear measurement noise is well
expressed as a Gaussian distribution, so the first term is
\begin{equation}
\label{pgt}
L(\bfk | {\bf m}) = (2\pi)^{-J/2} |{\bf N}|^{-1/2} \exp \left\{-{1 \over 2}
\left[
(\bfk - {\bf G}{\bf km})^T {\bf N}^{-1} (\bfk - {\bf G}{\bf km})
\right]\right\}.
\end{equation}
where ${\bf N}$ is the convergence noise matrix ${\rm diag}(\langle
\delta\kappa^2_s\rangle)$, and ${\bf k} = {\rm diag}(k_\ell)$.

For analytical convenience we take
the joint probability $L({\bf m},{\bf g})$ of the true and estimated
masses to be a bivariate Gaussian at each slice.  This is undoubtedly a poor
approximation, \eg\ most of the lensing cross-correlation power is on
angular scales where the mass distribution is highly non-linear and
non-Gaussian. We defer a more realistic treatment to later work. Given
our definition of the bias $B_i$ and correlation coefficient $r_i$, 
the probability can be written 
\begin{eqnarray}
\label{ptc}
L({\bf m}, {\bf g}) & = & (2\pi)^{-J} |{\bf C_u}|^{-1/2}
|{\bf C_g}|^{-1/2} \exp \left\{-{1 \over 2} \left[
({\bf m}-{\bf B}{\bf g})^T {\bf C_u}^{-1} ({\bf m}-{\bf B}{\bf g})
+ {\bf g}^T {\bf C_g}^{-1} {\bf g} \right]\right\} \\
{\bf C_g} & \equiv & {\rm diag}(\langle g_i^2 \rangle) \\
{\bf B} & \equiv & {\rm diag}(B_i) \\
{\bf C_u} & \equiv & {\rm diag}\left({B_i^2(1-r_i^2) \over r_i^2}
  \langle g_i^2 \rangle\right) = {\rm diag}\left[\langle m_i^2 \rangle
    (1-r_i^2)\right].
\end{eqnarray}
${\bf C_g}$ is the correlation matrix for the mass estimator, 
and ${\bf C_u}$ is the correlation matrix for the part of the mass
fluctuations that are {\em uncorrelated} with the estimator ${\bf
  g}$.  We assume the shells are thick enough for both matrices to be
diagonal. 

Multiplying Equations~(\ref{pgt}) and (\ref{ptc}), then
integrating over ${\bf m}$ gives the Gaussian distribution 
\begin{eqnarray}
\label{pgc}
L(\bfk, {\bf g}) & = & (2\pi)^{-J} |{\bf KC_g}|^{-1/2} \exp \left\{-{1 \over 2}
\left[ {\bf g}^T{\bf C_g}^{-1}{\bf g} + 
(\bfk - {\bf GkBg})^T {\bf K}^{-1} (\bfk - {\bf GkBg})
\right]\right\}, \\
\label{defk}
{\bf K} & = & {\bf N} + {\bf GkC_uk}^T{\bf G}^T.
\end{eqnarray}
${\bf K}$ is the covariance matrix for $\kappa$ if the mass estimators
$g$ are held fixed, \ie\ only the components of the mass distribution
that are uncorrelated with $g$ are considered stochastic.  

From this multivariate Gaussian distribution we may derive a Fisher
matrix using the formulae for zero-mean distribution given, \eg, by
\citet{TTH97}. We also multiply this by the number of independent lines
of sight in the survey to get
\begin{equation}
\label{fij}
F_{ij} = {4 \pi f_{\rm sky} \over \delta\Omega}
{\rm Tr}\left[ {\bf C_g}^{-1}{\bf C_g}_{,i}{\bf C_g}^{-1}{\bf C_g}_{,j}
 + {\bf K}^{-1}{\bf K}_{,i}{\bf K}^{-1}{\bf K}_{,j}
 + 2 {\bf K}^{-1}({\bf GkB})_{,i}{\bf C_g}({\bf
   GkB})^T_{,j}\right],
\end{equation}
with the commas in the subscripts denoting differentiation.  The
Fisher information nicely separates into three parts: the first is the
information that can be gleaned from the variances of the mass estimator
$g$, \ie\ the galaxy power spectrum.  The third term is Fisher
information that would arise from adjusting parameters to minimize the
$\chi^2$ in the fit of $\kappa$ to the estimated mass $g$ in
\eqq{gdiscrete}, were the values of $g$ taken as fixed and the matrix
${\bf K}$ taken as a known covariance for the $\kappa$ values.
The third term is information gleaned from the actual
covariance of the $\kappa$ residuals to this fit, and looks just like
the Fisher matrix for pure shear power-spectrum tomography, except that
the relevant mass power spectrum in this term is ${\bf C_u}$, the
power that is uncorrelated with the galaxies, not the (larger) total
power of $m$.  Hence the inclusion of cross-correlation will
lower the sample variance in lensing power-spectrum tomography.

In this paper we will presume that none of the (co-)variances of the
true or estimated mass distributions are well predicted by theory.
Hence we will marginalize over $C_g, B$, and $C_u$ at
each redshift.  There will be no cosmological
information gleaned
from analysis of ${\bf C_g}$ since we consider its elements to be free
parameters of the model, hence we can drop the first term of
\eqq{fij}.  

Since we are marginalizing over all the mass variances, we
can also absorb the overdensity-to-convergence conversion factor
$k_\ell$ into the definition of $m_\ell$ and $g_\ell$, and remove it
from Equations~(\ref{defk}) and (\ref{fij}).  We just must remember to
express our fiducial values of $C_u$ and $C_g$ as convergence
variances rather than as overdensity variances.

In a future paper we will consider cases where there are reliable
theoretical models for the mass fluctuations.  \citet{HJ04} consider
the case where not only the mass power spectrum but also the bias and
correlation of the galaxy sample are described by parametric models.

To recap, then, we will use the Fisher matrix
\begin{eqnarray}
\label{fij2}
F_{ij} & = & {4 \pi f_{\rm sky} \over \delta\Omega}
{\rm Tr}\left[ {\bf K}^{-1}{\bf K}_{,i}{\bf K}^{-1}{\bf K}_{,j}
 + 2 {\bf K}^{-1}({\bf GB})_{,i}{\bf C_g}({\bf
   GB})^T_{,j}\right], \\
{\bf K} & \equiv & {\bf N} + {\bf GC_u}{\bf G}^T.
\end{eqnarray}
with the parameters being:
\begin{itemize}
\item $\Omega_k$, which affects only ${\bf G}$.  The fiducial value is
  zero. 
\item The $d_i$, which also only affect ${\bf G}$.  The fiducial values
  at each redshift are taken from the $\Lambda$CDM model with
  $\Omega_m=0.3$. 
\item The diagonal elements of ${\bf C_u}$, over which we
  will marginalize to allow for our ignorance of the
  cross-correlations $r_\ell$ between the proxy and true mass
  distributions.  We take a fiducial $r_\ell=0.7$ but investigate the
  effect of different fiducial values.
\item The diagonal elements $B_i$ of ${\bf B}$, the bias factors, over
  which we will marginalize.  We take the
  fiducial $B_i=1$ without loss of generality.
\item The diagonal elements of ${\bf C_g}$ are also free
  parameters, but striking the first term of (\ref{fij}) is equivalent
  to marginalizing over these parameters.
\end{itemize}
With these parameters, all of the derivatives required in \eqq{fij2}
are simple, sparse matrices.

\subsection{Additional Information}
The CCC Fisher matrix will be singular due to the degeneracies in the
parameters noted above.  To this we must add a Fisher matrix for the
$d_i$ determined from the baryon acoustic oscillations; this
matrix will be
diagonal as the galaxy power spectrum measurements are independent
from shell to shell if the shells are much thicker than the acoustic horizon
scale. 

For the baryon acoustic scale measurement, we use an abstraction of
the full analysis provided by \citet{Seo}.  We presume that there is
a fractional uncertainty $\sigma_{\ln (1+z)}$ in $1+z$ on each
photometric redshift, which defines the depth of a ``slab'' of modes
in $k$-space that are measurable.  We also presume that the power
spectrum determination for all modes within the slab will be limited
by sample variance, which requires that the comoving volume density of
galaxies with good photo-z's must be $dN/dV \gtrsim 10^{-3} h^3\,{\rm
  Mpc}^{-3}=10^{7.5}(H_0/c)^3$ to make the shot noise negligible at the highest
$k\approx0.2h\,{\rm Mpc}^{-1}$ where baryon oscillations are usefully
detected.  When these conditions are met, the fractional error in the
mean $d$ for a given shell will scale as $\sqrt{\sigma_{\ln
  (1+z)}/V}$.  The prefactor depends upon the intensity of the wiggles
in $P(k)$, which diminishes somewhat at lower redshift as nonlinear
effects erase the acoustic features.  We approximate this by assigning
an uncertainty
\begin{equation}
\label{baoerrors}
{\sigma_d \over d} = 0.005 \left( {1 + z \over 2 + z}\right) (VH_0^3/c^3)^{-1/2}
\left({\sigma_{\ln (1+z)} \over 0.04}\right)^{1/2}
\end{equation}
for those shells which, by \eqq{dndz}, have $dN/dV$ above the
sample-variance-limited threshold noted above.  If the galaxy density
in the shell is below this level, we assume no constraint on $d$ from
BAO.  Our simplification fits the \citet{Seo} results well and comes
within 30\% of the forecasts by \citet{Glazebrook}.

We will also investigate the impact of a regularization constraint on
the $B$ function.  Writing the bias as a function of the lens-plane
angular diameter distance $\ell$, we add to the likelihood a
function
\begin{equation}
\label{breg}
-\ln L = {1 \over R^2\ell_{\rm max}} \int_0^{\ell_{\rm max}} d\ell\,
\left({dB \over d\ell}\right)^2.
\end{equation}
The parameter $R$ specifies a scale of the RMS bias slope that will be
suppressed.  The discretized version of this likelihood is a quadratic
form in the $B_i$ and easily incorporated into the
Fisher matrix.
We project the three degeneracies in
Equations~(\ref{degen1}--\ref{degen3}) out of the ${\bf B}$ vector
before applying the regularization constraint, so that the
regularization does not artificially break these degeneracies.

\section{Curvature Forecasts}
We calculate the Fisher matrix about our fiducial cosmology using the
following estimates for the relevant quantities.

The $\kappa$ variances in ${\bf N}$ are taken from
\eqq{vargamma} with the source galaxies distributed as
\begin{equation}
\label{dndz}
{dn \over dz } \propto z^2 \exp\left[-(z/1.41z_0)^{1.5}\right],
\end{equation}
with the median redshift $z_0$ and the integrated density $n$ chosen to
crudely mimic the
expectations of future surveys.  Our default is to take
$z_0=1.5$, $n=100\,{\rm arcmin}^{-2}$, the performance one might
expect from a space-based survey like the proposed SNAP wide survey
\citep{paper2,wlerrs}. 

We set the variance $\langle m^2_\ell \rangle$ to
the convergence variance $d\langle \kappa^2 \rangle / dz$ integrated over
each shell, using the nonlinear dark-matter power spectrum for a
$\Lambda$CDM model, restricted to $k<2\pi/(50~{\rm kpc})$, as detailed in
BJ04.  Given a fiducial correlation coefficient $r$, we set 
$\langle g_\ell^2\rangle=r^2\langle m_\ell^2 \rangle$.  We will set
all the fiducial $r_\ell=0.7$, although the marginalization allows
them to have independent solutions. 

We also truncate the source and lens distributions at $z_{\rm max}=3$
unless otherwise noted.  The number of redshift bins will range from
30 to $\approx200$ to search for effects of the artificial
discretization of the problem.

\subsection{Constraints on $\Omega_k$}
The Fisher matrix for the fiducial cosmology yields, first, an
estimate of the uncertainty on $\Omega_k$ of
$\sigma_k=1.6\times10^{-4} f_{\rm sky}^{-1/2}$ in the unrealistic case
where all the 
biases, distances, and correlation coefficients are known {\it a
  priori}.  This is in good agreement with our crude guess earlier.

When we marginalize over all parameters except $\Omega_k$, we find
that {\em if we project out the degeneracy in Equation
  (\ref{degen3})}, then
the weak lensing CCC method is remarkably efficient at extracting
$\Omega_k$, giving an uncertainty $\sigma_k\approx 2\times10^{-4}
f_{\rm sky}^{-1/2}$, nearly as good as before any marginalization.
Of course this is akin to stating that the Titanic was watertight,
aside from the big gash from the iceberg.  Nonetheless this feature of the
CCC method is illuminating: it tells us that the role of other
distance measures, \ie\ supernovae and transverse BAO,
will be purely to bound the size of the quadratic
degeneracy in the CCC distances that couples to $\Omega_k$, because
the CCC method reduces any {\em other} form of uncertainty in $d(z)$ and
$\Omega_k$ to far below the level achievable from such experiments.

If, for example, we remove the three singular values from  the WL-only
Fisher matrix, and then estimate the uncertainties in $d(z)$, we find
$\sigma_d\approx1.4\times10^{-4}f_{\rm sky}^{-1/2}$ over the range
$0.6\lesssim z 
\lesssim 2.5$, when the distances are averaged over
$\Delta z = 0.1$ bins.  This is 1--2 orders of magnitude better than
the most ambitious targets for supernova measurements, and
$\sim10\times$  better
than a spectroscopic BAO survey of equivalent sky coverage.

The final constraint on $\Omega_k$, therefore, comes down to the
simple question of how well the non-WL data can constrain a variation
of the form
\begin{equation}
d \rightarrow d(1 + \alpha_0 + \alpha_1d + \Omega_kd^2/2),
\end{equation}
with the WL CCC data having locked down any other form of
variation.  Our simplified analysis of the transverse BAO survey
available from the canonical WL survey data yields
\begin{equation}
\label{sigk}
\sigma_k = 0.04 f_{\rm sky}^{-1/2} \left({\sigma_{\ln z} \over
  0.04}\right)^{1/2}.
\end{equation}

\begin{figure}[t]
\epsscale{0.7}
\plotone{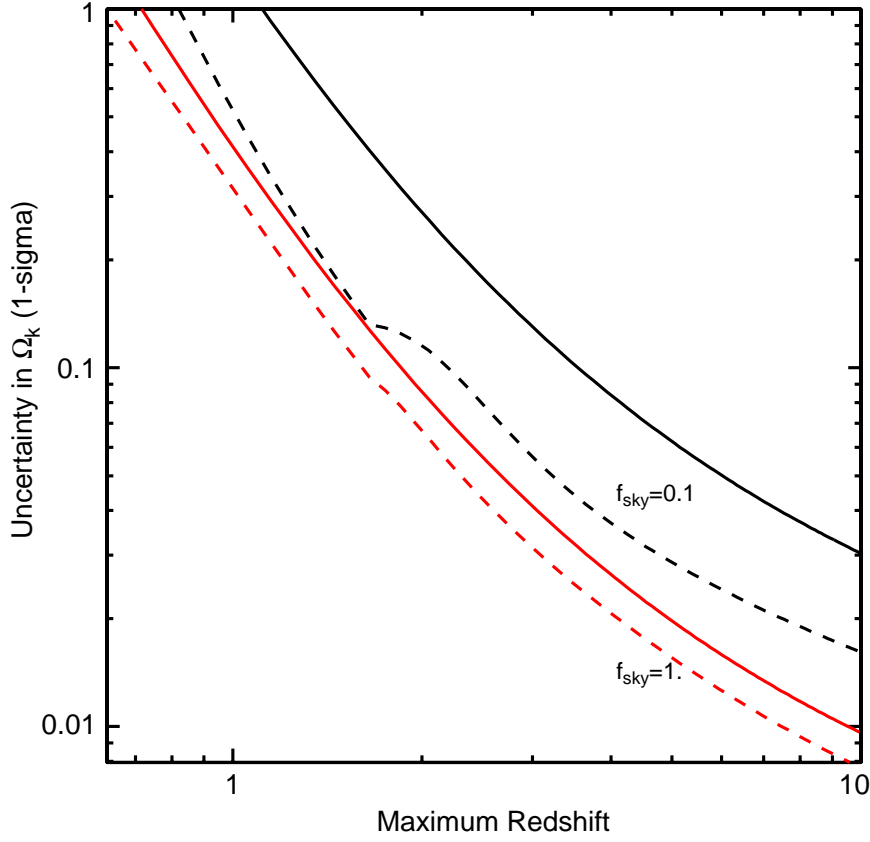}
\caption[]{
\small Uncertainties obtained on $\Omega_k$ are plotted vs the maximum
redshift for which weak lensing cross-correlation and baryon acoustic
oscillation data are {\em jointly} obtained.  These are purely metric
constraints on $\Omega_k$, and it is presumed that the
WL data constrain all forms of variation of $\Omega_k$ except the
principle degeneracy described in the text.  The upper two curves are
for 10\% sky coverage, the lower two for full-sky.  In each pair
the solid line is for WL$+$BAO constraints, and the dashed line shows
the effect of adding 1\% measures of supernova distances in each
$\Delta z=0.1$ bin for $0.4<z<1.7$.
}
\label{klims}
\end{figure}

In our canonical survey, both WL information and
sample-variance-limited transverse BAO data are available over the
range $0<z<3$.  In Figure~\ref{klims} we examine the dependence of the
$\Omega_k$ constraint on the redshift range over which these data are
jointly available.  Reaching a 1\% model-independent constraint on
$\Omega_k$ would require a massive spectroscopic BAO survey, and/or
obtaining CCC$+$BAO data to $z\sim8$.  This is not a 
completely ridiculous prospect: the CMB anisotropy is, after all, a
transverse BAO measurement at $z=1100$.  Lensing of the CMB is also
observable, but even more lensing information may be available using
the 21-cm emission from neutral hydrogen in the early stages of
reionization \citep{Pen}.

Figure~\ref{klims} also plots the $\Omega_k$ limits that result from
combining the transverse BAO distance measurements with a supernova survey that
yields independent 1\% measurements of $d(z)$ at $\Delta z=0.1$ steps
in the range $0.4\le 1.7$.  We find that the SN data can cut
$\sigma_k$ in half for a transverse BAO survey with $f_{\rm sky}=0.1$, but the
SN data are of only slight help (20\%) to a full-sky photo-$z$ BAO
survey.  

We note that the Fisher uncertainties on $\Omega_k$ are quite
independent of the number of redshift shells or of any regularization
constraint on the bias factor.  They are also essentially independent
of the assumed proxy correlation coefficient $r$ or the WL shape noise
$\sigma_\gamma$, because the nominal WL survey is already reducing the
errors to insignificance in any mode for which it has power at all.
The galaxy distribution parameters $z_0$ and $n$ are important only
in that they have been used to determine the redshift range over which
sample-variance-limited BAO data are available.  If we scale back to
$z_0=1.0$, $n=40\,{\rm arcmin}^{-2}$, as might be expected for a
deep ground-based survey \citep{wlerrs}, then the uncertainty in
$\Omega_k$ triples to $0.12f_{\rm sky}^{-1/2}$ for the nominal
$\sigma_{\ln z}=0.04$ because the galaxy density drops below the BAO
sample-variance limit at $z<3$.  In a real survey, the distribution of galaxies
with accurately measured shapes will not, however, equal the
distribution of those with accurately measured photometric redshifts,
so a more detailed analysis is required in order to compare surveys.
It is clear, however, that sky coverage, photo-$z$ accuracy, and
redshift range of {\em joint} CCC-BAO data are the key characteristics
for this geometric constraint of $\Omega_k$.

\subsection{Effects of Systematic Errors}
The WL CCC method depends upon measuring shear to very high accuracy.
Indeed the cosmology dependence of shear is so subtle that
\citet{Mandelbaum} reverse the technique, and use the redshift
dependence of galaxy-shear correlations as a 
test for systematic errors in shear measurement.
Any such shear systematics could substantially
degrade the curvature constraints derived above.  But we show here
that the CCC data will be rich enough to solve simultaneously for
cosmology and calibration systematics.  

The WL
cross-correlation technique is insensitive to spurious shear
signals caused by PSF ellipticities, because they will not correlate
with the proxy mass fields.  Calibration errors on the shear will,
however, be important.
We examine a model,
similar to that of \citet{Ishak04}, in which the shear on source shell
$s$ is mismeasured by a factor $(1+f_s)$.  This is simply incorporated
into our governing equations by setting
\begin{equation}
G_{s \ell} = (1+f_s) {d_s - d_\ell \over d_s}(1-\Omega_kd_sd_\ell/2)
\end{equation}
for $d_s>d_\ell$.  The Fisher matrix now has the parameters $f_i$ as
well.  We assume a Gaussian prior distribution on the $f_i$ for which
they are all independent, with ${\rm Var} f_i = \sigma^2_f$.

We find the constraints on $\Omega_k$ from the canonical survey are
essentially unchanged if $\sigma_f\lesssim 0.01 f_{\rm sky}^{-1/2}$, a
level that is probably achievable from ground or space observations.
For $\sigma_f=1$, 
\ie\ complete ignorance of the calibration, $\sigma_k$ rises to 0.067
for the full-sky survey.  The degradation in the curvature constraint
is mild because even a ``self-calibrated'' CCC survey
constrains the non-degenerate modes fairly well, and the nature of the
degeneracies is only slightly expanded by the additional free
parameters.  A strictly $z_s$-dependent calibration error does not
appear to be a problem for the curvature measurement.

\section{Parametric Modelling}
The Fisher matrix constructed over parameters $d_i$ and $\Omega_k$,
marginalized over the nuisance parameters, fully describes the
cosmological information content of a CCC survey.  Distance
information from transverse BAO and supernovae
can be summed into the same matrix.  If we now consider each $d_i$ to
correspond to a known redshift $z_i$ (which was unimportant to the
metric curvature determination), this Fisher matrix describes our
knowledge of the expansion history, and can be used
to forecast constraints on parametric models of
the expansion of the Universe.  To project the $d$--$\Omega_k$ Fisher matrix
onto some parameters $p_i$, we simply need to know $dD_A(z)/dp_i$.
Calibration errors on the shear can be included as described above;
biases $\Delta z_p(z)$ in the photometric redshift scale are also
easily included as they give rise to perturbations in the measured
distances equal to $[d(d_i)/dz]\,\Delta z_p(z)$. 

The distance-based analysis is also useful in that it has revealed the
nature of the constraints from the various methods.  The WL
cross-correlations are completely blind to changes in $\log d$ that
are quadratic (linear) in $d$ when $\Omega_k$ is free (fixed), but
offer very strong limits on any other forms of variation in $d(z)$.
When WL is used in combination with transverse BAO and/or supernova
measurements, the latter will serve primarily to constrain these low-order
degeneracies. 

\begin{figure}[t]
\plotone{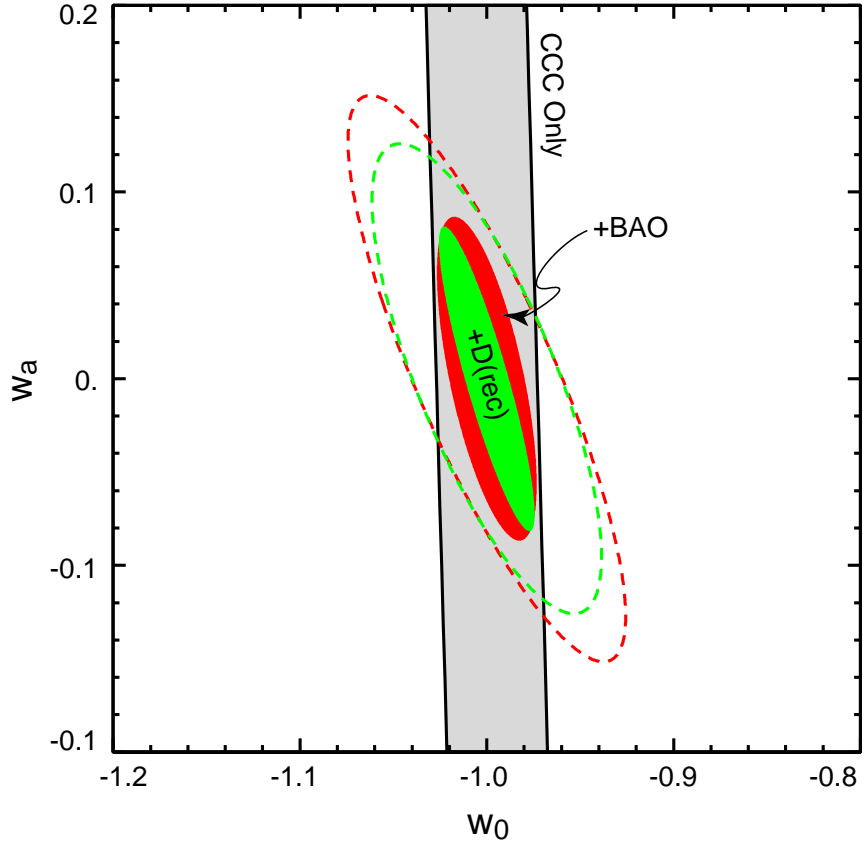}
\caption[]{
\small Error ellipses in the $w_0$--$w_a$ plane for parametric models
of dark energy that have been marginalized over $\Omega_m$ and
$\Omega_k$.  These are 68\% CL regions ($\Delta\chi^2=2.3$), for
full-sky coverage; errors scale as $f_{\rm sky}^{-1/2}$.  The
large ellipse is for WL CCC information only.  The middle shaded
ellipse (red) adds transverse baryon acoustic oscillation data from the
photo-$z$ information of the WL survey.  The innermost (green) shaded
region includes a measurement of the distance to recombination from
expected CMB data.  The two dashed ellipses allow for systematic shear
calibration errors of 3\% per redshift bin, which is in the
``self-calibration'' regime.  The outer ellipse is CCC$+$BAO, the
inner adds the recombination distance.
}
\label{ellipses}
\end{figure}

As an example we consider a model for the expansion in which the free
parameters are $\Omega_k$, $\Omega_m$, and the 
parameters $w_0$ and $w_a$ for the dark energy equation of state $w=w_0 +
w_a(1-a)$.  Note that $\Omega_{\rm DE}=1-\Omega_m-\Omega_k$. 
We will also quote the uncertainty in $w_p$, the equation of state at
the ``pivot redshift'' $z_p=1/a_p-1$ for which the model $w=w_p+(a_p-a)w_a$
has uncorrelated Fisher errors in $w_p$ and $w_a$.
We consider the constraints that arise from our canonical
survey of weak lensing cross-correlations and photo-$z$ BAO data.  We
presume a fiducial proxy-mass correlation coefficient of $r=0.7$.
Table~\ref{sigtable} lists the Fisher-matrix estimates of the standard
deviations 
on these four parameters.  Figure~\ref{ellipses} plots error ellipses
in the $w_0$--$w_a$ plane after marginalization over the other two
variables. 

\begin{deluxetable}{cccccc}
\tablewidth{0pt}
\tablecaption{CCC$+$BAO Constraints on Cosmological Parameters}
\tablehead{
\colhead{Photo-$z$ Error} &
\colhead{Calibration Prior\tablenotemark{a}} &
\multicolumn{4}{c}{Parameter Errors\tablenotemark{a}} \\
\colhead{$\sigma_{\ln z}$} &
\colhead{$\sigma_f$} &
\colhead{$\sigma(\Omega_k)$} &
\colhead{$\sigma(\Omega_m)$} &
\colhead{$\sigma(w_p)$} &
\colhead{$\sigma(w_a)$}}
\startdata
CCC only & \nodata & 0.36 & 0.60 & 0.018 & 2.0 \\
CCC only & \nodata & \nodata & 0.015 & 0.017 & 0.081 \\
0.04 & \nodata & 0.0066 & 0.0045 & 0.013 & 0.057 \\
0.04 & \nodata & \nodata & 0.0038 & 0.010 & 0.052 \\
0.04 & 0.03 & 0.013 & 0.007 & 0.044 & 0.100 \\
0.04 & 0.03 & \nodata & 0.007 & 0.026 & 0.083 
\enddata
\tablenotetext{a}{All uncertainties are for the fiducial 
  survey described in the text, with full-sky coverage,
  but scale with $f_{\rm sky}^{-1/2}$ as long as the calibration prior
  is similarly scaled.  Dashes indicate quantities that are held
  fixed. Each error assumes marginalization over all the other
  parameters.} 
\label{sigtable}
\end{deluxetable}

We find first that the WL CCC alone retains a strong degeneracy
between $\Omega_k$ and the other parameters, with large errors of
0.36, 0.60, and 2.0 on $\Omega_k$, $\Omega_m$, and $w_a$,
respectively, but with correlation coefficients $>0.99$ among all of
these.  [These are for full-sky surveys; we will omit the factor
$f_{\rm sky}^{-1/2}$ for brevity here.]  The degeneracy is broken
either by assuming a flat Universe, by using the CMB distance to
recombination, or by incorporating the transverse
BAO information over the same sky fraction.  In the last case,
$\Omega_k$ has uncertainty 0.007, which as expected is a much tighter
constraint than the model-independent one above.  
We also note that
the $w_p, w_a$ uncertainties of (0.013, 0.057) are only slightly
reduced by assuming flatness, although their $z_p$ shifts.

Inclusion of an accurate (0.1\%) angular-diameter distance to
recombination from the CMB, as might be expected from Planck, drops
uncertainties on $\Omega_k$ by a factor 3, and a factor 2 for $w_p$.

\subsection{Scaling with Survey Characteristics}
The BJ04 analysis asserts that CCC constraints improve without bound
as $n^{-1/2}$.  ZHS did not concur, and we would now
agree.  Since Stebbins (private communication) proves that
${\bf G}$ must be non-singular in the continuum limit (and we have
constructed its inverse), 
then the matrix ${\bf K}={\bf N} +{\bf GC_uG}^T$ will remain
non-singular as the shape noise ${\bf N}\rightarrow 0$, meaning that
the Fisher information in \eqq{fij} remains finite.  As a practical
matter, we find that
in the range of $n$ accessible to planned observations, CCC Fisher
information scales only slightly more slowly than $n^{-1/2}$.  There
is no simple scaling once we combine the CCC information with that
from BAO.  The constraints are nearly independent of the choice of
$\delta\Omega$, again because the shape noise dominates sample
variance. 

The BJ04 derivation also leads to CCC-only parametric uncertainties that
scale nearly inversely with the fiducial correlation coefficient
$r$.  In the present analysis we see that such scaling arises if we
consider only the effect of $r$ upon the ${\bf C_g}$ factor of
\eqq{fij2}, and if we drop the left-hand term entirely.  But if the
effect of $r$ upon the ${\bf C_u}$ terms in the noise matrix ${\bf K}$
are also considered, then the dependence is more complex.
We find that the CCC-only parametric
constraints on flat Universes scale more weakly than inversely with
$r$ over the 
range 0.3--1.0.  The dependence is yet weaker once we include the BAO
information. 

\subsection{Effects of Calibration Errors}
\begin{figure}[t]
\plotone{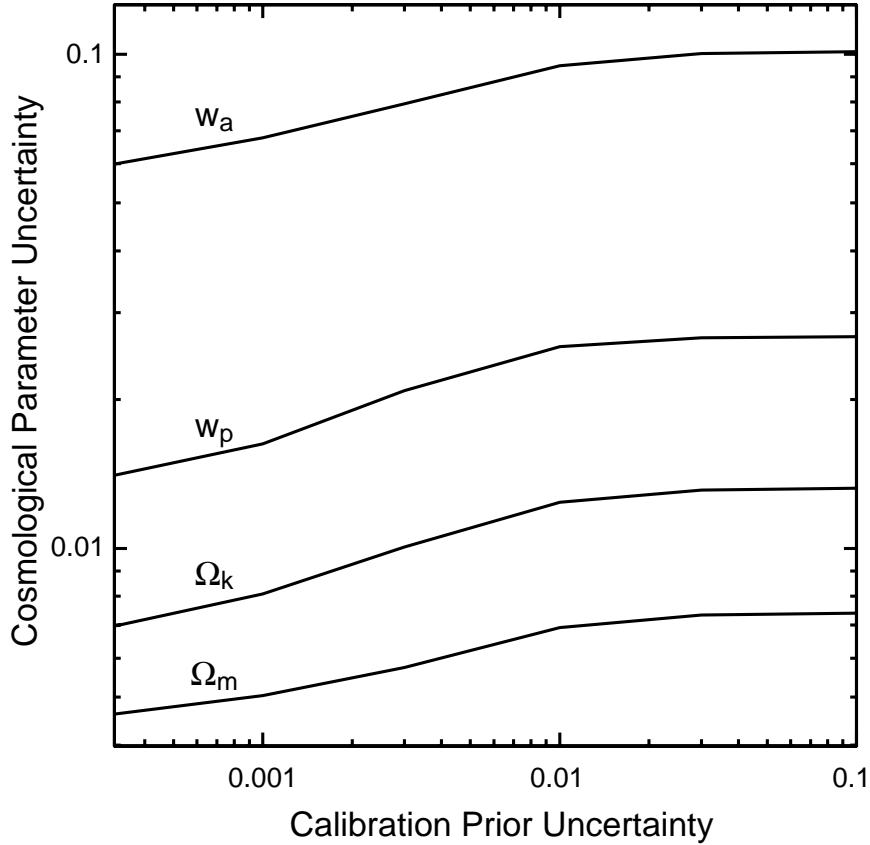}
\caption[]{
\small The 1-$\sigma$ uncertainties in each of the labelled
cosmological parameters in a CCC$+$BAO survey are plotted vs the
accuracy of our prior knowledge of the shear calibration factor $f_i$
in each redshift bin.  We see that calibration errors smaller than
$\approx10^{-3}f_{\rm sky}^{-1/2}$ are harmless, and there is a
``self-calibration'' regime for $\sigma_f>0.01f_{\rm sky}^{-1/2}$, in
which the CCC data solve for the calibration factors to an accuracy
that is better than the prior.  The degradation from self-calibration
is mild.  Both axes may be scaled jointly with $f_{\rm sky}^{-1/2}$.
}
\label{fcal}
\end{figure}

Figure~\ref{fcal} plots the degradation of the parametric cosmology
constraints as we weaken the prior on the calibration factors $f_i$.
There are three regimes: when the calibration is known {\it
  a priori} to better than $\approx10^{-3}f_{\rm sky}^{-1/2}$, there
is no significant degradation due to calibration errors.  Over the
range $10^{-3}f_{\rm sky}^{-1/2}<\sigma_f<0.01f_{\rm sky}^{-1/2}$, the
constraints continually degrade.  Calibration uncertainties above
$\approx0.01f_{\rm sky}^{-1/2}$, however, cause no further degradation
in cosmological constraints, as the CCC data solve for the systematic
factors in a ``self-calibration'' regime.  The self-calibration regime
has parameter errors that are 1.6--2 times larger than the
no-systematic regime.  We conclude that lowering the calibration
errors from 0.01 to 0.001 is equivalent to increasing the survey area
by a factor of 2.5--4, but interesting cosmological constraints are
possible even with poor {\it a priori} calibration systematics.

We find a similar scaling of uncertainties with calibration
systematics for a shallower, ground-based survey.

\subsection{Comparison to Previous Work}
We alter the parameters of our fiducial survey to match those of
ZHS, namely we set $f_{\rm sky}=0.1$, $z_0=1.0$,
$\sigma_\gamma=0.3/\sqrt2$, $\Omega_k=0$, and we use an equation of
state $w=w_0+w^\prime z$ instead of the $w_a$ form.  The ZHS
analysis also allows for an arbitrary distribution of mass redshift
within a shell, and accounts for the signal attenuation due to
photo-$z$ errors.  We compare to their $\sigma_z=0.01$ case.  For a
fixed-$w$ model with only CCC information, we obtain standard
deviations on $(\Omega_m, w)$ of (0.028, 0.046) for $r=0.7$, roughly
$2\times$ smaller than ZHS derive.  We must lower $r$ to $0.25$ to
obtain similar results.  They also derive errors on $(w_0, w^\prime)$
when marginalized over $\Omega_m$ with a prior of 0.03.  In this case
our $r=0.7$ constraints of (0.03,0.11) compare to their
(0.07,0.09)---similar in size but different in shape.

The agreement to within factor two must be considered adequate at this
time.  While there remain substantial differences in the analytical
approach to the CCC method, the discrepancies could easily be
attributed to different assumptions about the correlation between the
mass and the proxy field.  ZHS, as well as \citet{HJ04}, 
take the galaxy distribution itself as the proxy field, and assume
that the galaxies populate the mass via a pure Poisson process.  In
this case the correlation coefficient between 
multipole $l$ of the proxy and mass fields at distance $D_A$ is
\begin{equation}
r = \left\{ 1 + \left[{dN \over dV}
  P(l/D_A)\right]^{-1}\right\}^{-1/2}, 
\end{equation}
where $P(k)$ is the 3-dimensional mass power spectrum.  Thus when
$nP<1$, $r$ drops below the 0.7 value of our fiducial model.  This
occurs at $l\gtrsim10^4$ in our fiducial model for $0.5<z<2$,
whereas our fiducial model makes use of shear power to higher $l$.
Were we to adopt a Poisson model our parametric dark energy
constraints would weaken.  It is not clear, however, that the Poisson
model is appropriate once we approach the length scales of group- and
galaxy-scale halos, which are currently suspected of harboring a
single central galaxy with near-unity probability, plus satellites.
The galaxy--mass 
covariance on small scales may thus be a good deal stronger than
Poisson models predict.  This is surely a subject for further
research. 

We may also compare to the CCC-only results with the BJ04 method by
reverting to the parameters assumed therein for the ``SNAP'' case:
$f_{\rm sky}=0.025$, $r=0.8$, $\sigma_\gamma=0.15$, $\Omega_k=0$, and
a prior of 0.03 on $\Omega_m$.  We find a pivot redshift $z_p$ and
uncertainty $\sigma(w_p)$ that are the same in both cases, but $w_a$
is 35\% lower with the present Fisher matrix.  This could be a result
of some of the simplifications made by BJ04 in construction of the
covariance matrix for the $X_{s\ell}$ in \eqq{expect}.  Alternatively
it could derive from our present assumption that the convergence, mass, and
proxy field values have a multivariate Gaussian distribution, while
BJ04 have the 
less restrictive assumption that the $X_{s \ell}$ have a
Gaussian distribution.

A close comparison with \citet{HJ04} is well beyond our present scope,
because these authors include a parametric model for the bias function
and limit the cross-correlation information to multipoles
$\ell<3000$.  In principle both differences can be accommodated in the
present formalism, by switching to a spherical-harmonic decomposition,
and by projecting parametric bias constraints onto
the $\Omega_k$--$d_i$--$B_i$ Fisher matrix before marginalizing over
the $B_i$.  A WL CCC survey reduces the errors on $\ln B$
to $\lesssim 10^{-3} f_{\rm sky}^{-1/2}$, apart from a quadratic
degeneracy with distance.  Hence the {\it a priori} theoretical
constraints would need to be stronger than this in order to improve
dark energy measurements.  We also note that Hu \& Jain take a
Poisson model for the mass-galaxy correlation, and assume a density of
lens galaxy samples that is 2 orders of magnitude lower than assumed
here, which we believe accounts for the cross-correlation constraints
being much weaker in that work than in the present calculations.

A short summary of this subsection is that there are currently
multiple formalisms for analysis of CCC data, and they currently
differ by up to a factor 2 in Fisher uncertainties on parametric
dark-energy models.  Further work is certainly required in order to
understand the nature of the CCC constraints, and to understand our
ability to reconstruct the mass distribution from the galaxy data.

\section{Discussion}
Reparameterizing the observable galaxies by their (dimensionless)
angular diameter distance $d$ rather than redshift $z$ makes it clear
how the WL CCC method can provide model-independent geometric constraints on
$\Omega_k$.  This methodology also reveals that CCC data is degenerate
under alterations to $\ln b$ and $\ln d$ by quadratic functions of
$d$, if $\Omega_k$ is free. 
Apart from three exact degeneracies, the cross-correlation
strength, being a joint 
function of the lens and source distances, is very efficient at
producing decoupled
estimates of $\Omega_k$, the distances $d_i$
and the bias factors $B_i$, and can even determine shear calibration
factors $f_i$ for each source plane with modest degradations of factor
$\lesssim 2$ in cosmological accuracy.  

With sufficiently accurate photometric redshifts, the same survey data
used for weak lensing may be used to determine the transverse BAO
scale.  This gives additional
model-independent constraints on $d$ that are of lower precision than
the CCC data, but are completely free of degeneracy.  Such a combined
CCC-BAO survey should yield uncertainties of $\approx0.04 f_{\rm
  sky}^{-1/2}$ on $\Omega_k$.  We reiterate that such constraints are
completely independent of any assumptions about the matter-energy
content of the Universe, any biases in photometric redshifts, or in
fact any alterations to the Friedmann equations or the deflection
equations for light.  They merely require that the Robertson-Walker
metric be applicable to our Universe.  Given the lack of viable dark-energy
theories, it seems prudent to seek cosmological information that
remains valid even if the acceleration is attributable to an
alteration of General Relativity rather than a previously unnoticed
stress-energy contribution.

In the spirit of conservatism, we should note that the RW metric may
not be sufficiently accurate.  The clumpiness of the matter-energy
distribution may invalidate our adoption of the filled-beam angular
diameter distance, and a more sophisticated treatment may be required
\citep{HL}.  It is not clear to what extent small-scale clumpiness
can influence the apparent shear of galaxy-scale images as surveyed
across the entire sky---this is quite a different regime from studies
of strongly lensed quasars or supernovae for which the Dyer-Roeder
distances have been most carefully studied.

The curvature measurement is limited by the ability to break the WL
CCC degeneracy, \ie\ by the accuracy and redshift span of the BAO or
SN measurements of angular-diameter distances.  Our baseline
constraint assumes photo-$z$ BAO information for $0<z<3$ over the full sky.  
A massive spectroscopic BAO survey
would offer substantial improvement, but Type Ia supernova
studies would not offer substantial improvement over full-sky
photo-$z$ BAO unless systematic uncertainties in SNIa peak magnitudes
could be brought well below 1\% and the redshift range extended beyond
$z=2$. 
Inclusion of lensing and acoustic-scale information from the epochs of
recombination or reionization would help significantly.

The parametric dark-energy constraints derived from the CCC$+$BAO
information are more fluid, as our derivation has made somewhat
arbitrary assumptions about the correlation between the true mass and
the proxy mass derived from the galaxy distribution.  The simplifying
assumptions about Gaussianity, uncorrelated mass shells, and
uncorrelated lines of sight should
also be improved upon.  The spherical-harmonic-based Fisher analyses
of \citet{HJ04}, \citet{SK04}, and ZHS are in this respect
superior to our treatment of uncorrelated lines of sight.  We note
however that the Fisher matrix from \eqq{fij} is applicable to
multipole coefficients.
More work is also required to examine the impact of finite photo-$z$
precision on the CCC constraints, as per ZHS.

A weak lensing survey would produce other kinds of information which
we have not used in this analysis.  Power-spectrum \citep{Hu,paper3} and
bispectrum \citep{TJ04} tomography, and WL cluster counts
\citep{WK03,Marian} constrain dark energy once a model
for the growth of structure $G(z)$ is specified.  One could augment
our current $\Omega_k-d_i$ Fisher matrix with growth factors $G(d_i)$
to produce a non-parametric information estimate, which could then be
projected onto the parameters of specific models, even those which
predict changes to the usual equation for linear growth of
perturbations in an expanding Universe.  

The method proposed herein for metric constraints on curvature, and
for strong parametric constraints on cosmology, will require
substantial practical advances before implementation, plus additional
theoretical work.  Progress in high-accuracy shape and photo-$z$
measurement is needed, as is research into intrinsic galaxy-shape
correlations \citep{Hirata2}, non-linearities in lensing
\citep{White}, and methods of reconstructing the matter distribution
from a galaxy distribution.  Nonetheless the WL CCC
method shares with the BAO method the ability to generate extremely
robust and model-independent cosmological constraints, particularly
when they are used in combination.

\acknowledgements
This work is supported in this work by grant AST-0236702 from the
National Science Foundation, and Department of Energy grant
DOE-DE-FG02-95ER40893.
This work was motivated by the rants of Eric Linder, Dan Eisenstein,
and Ned Wright on the nature of current curvature constraints.  I
thank Bhuvnesh Jain, Jacek Guzik, Albert Stebbins, Eric Linder, and
Dan Eisenstein for many helpful conversations, and Jun Zhang for
providing numerical results from his work.

\end{document}